\documentclass[aps,prl,twocolumn,superscriptaddress,showpacs,amsmath]{revtex4}
\usepackage{graphicx}
\usepackage{bm}

\begin{document}

\title{Large enhancement of radiative strength for soft transitions in the 
quasicontinuum}

\author{A.~Voinov}
\email{voinov@nf.jinr.ru}
\affiliation{Frank Laboratory of Neutron Physics, Joint Institute of Nuclear 
Research, 141980 Dubna, Moscow region, Russia}
\author{E.~Algin}
\affiliation{Lawrence Livermore National Laboratory, L-414, 7000 East Avenue, 
Livermore, California 94551}
\affiliation{North Carolina State University, Raleigh, North Carolina 27695}
\affiliation{Triangle Universities Nuclear Laboratory, Durham, North Carolina 
27708}
\affiliation{Department of Physics, Osmangazi University, Meselik, Eskisehir, 
26480 Turkey}
\author{U.~Agvaanluvsan}
\affiliation{Lawrence Livermore National Laboratory, L-414, 7000 East Avenue, 
Livermore, California 94551}
\affiliation{North Carolina State University, Raleigh, North Carolina 27695}
\affiliation{Triangle Universities Nuclear Laboratory, Durham, North Carolina 
27708}
\author{T.~Belgya}
\affiliation{Institute of Isotope and Surface Chemistry, Chemical Research 
Centre HAS, P.O.Box 77, H-1525 Budapest, Hungary}
\author{R.~Chankova}
\affiliation{Department of Physics, University of Oslo, N-0316 Oslo, Norway}
\author{M.~Guttormsen}
\affiliation{Department of Physics, University of Oslo, N-0316 Oslo, Norway}
\author{G.E.~Mitchell}
\affiliation{North Carolina State University, Raleigh, North Carolina 27695}
\affiliation{Triangle Universities Nuclear Laboratory, Durham, North Carolina 
27708}
\author{J.~Rekstad}
\affiliation{Department of Physics, University of Oslo, N-0316 Oslo, Norway}
\author{A.~Schiller}
\email{schiller@nscl.msu.edu}
\affiliation{Lawrence Livermore National Laboratory, L-414, 7000 East Avenue, 
Livermore, California 94551}
\author{S.~Siem}
\affiliation{Department of Physics, University of Oslo, N-0316 Oslo, Norway}

\begin{abstract}
Radiative strength functions (RSFs) for the $^{56,57}$Fe nuclei below the 
separation energy are obtained from the $^{57}$Fe$(^3$He$,\alpha\gamma)^{56}$Fe
and $^{57}$Fe$(^3$He$,^3$He$^\prime\gamma)^{57}$Fe reactions, respectively. 
An enhancement of more than a factor of ten over common theoretical models of 
the soft ($E_\gamma\alt 2$~MeV) RSF for transitions in the quasicontinuum 
(several MeV above the yrast line) is observed. Two-step cascade intensities 
with soft primary transitions from the $^{56}$Fe$(n,2\gamma)^{57}$Fe reaction 
confirm the enhancement.
\end{abstract}

\pacs{25.40.Lw, 25.55.Hp, 25.20.Lj, 27.40.+z}

\maketitle

Unresolved transitions in the nuclear $\gamma$-ray cascade produced in the 
decay of excited nuclei are best described by statistical concepts: a radiative
strength function (RSF) $f_{XL}(E_\gamma)$ for a transition with multipolarity 
$XL$ and energy $E_\gamma$, and a level density $\rho(E_i,J_i^\pi)$ for initial
states $i$ at energy $E_i$ with equal spin and parity $J_i^\pi$ yield the mean 
value of the partial decay width to a given final state $f$ \cite{BE73}
\begin{equation}
\Gamma_{if}^{XL}(E_\gamma)=f_{XL}(E_\gamma)\,E_\gamma^{2L+1}/\rho(E_i,J_i^\pi).
\label{eq:partial}
\end{equation}
Most information about the RSF has been obtained from photon-absorption 
experiments in the energy interval 8--20~MeV, i.e., for excitations above the 
neutron separation energy $S_n$. There, the giant electric dipole resonance 
(GEDR) is dominant. Data on the soft ($E_\gamma<3$--4~MeV) RSF for transitions 
in the quasicontinuum (several MeV above the yrast line) remain elusive. 
Corresponding data from discrete transitions show large fluctuations and are 
biased toward high transition strengths due to experimental thresholds. First 
data in the statistical regime have been obtained from the 
$^{147}$Sm$(n,\gamma\alpha)^{144}$Nd reaction \cite{Po82}. They indicate a 
moderate enhancement of the soft $E1$ RSF compared to a Lorentzian 
extrapolation of the GEDR\@. For spherical nuclei, in the framework of 
Fermi-liquid theory, this enhancement is explained by a temperature dependence 
of the GEDR width \cite{KM83}, the Kadmenski{\u{\i}}-Markushev-Furman (KMF) 
model. However, the experimental technique requires the presence of 
sufficiently large $\alpha$ widths and depends on estimates of both $\alpha$ 
and total radiative widths in the quasicontinuum below $S_n$.

The sequential extraction method developed at the Oslo Cyclotron Laboratory 
(OCL) \cite{SB00} has enabled further investigations of the soft RSF by 
providing unique data for transitions in the quasicontinuum with sufficient 
averaging. For deformed rare-earth nuclei, it has been shown that the RSF can 
be described in terms of a KMF GEDR model, a spin-flip giant magnetic dipole 
resonance (GMDR), and a soft $M1$ resonance \cite{VG01,SV04}. In this work, we 
report on the first observation of a strong enhancement of the soft RSF in 
$^{56,57}$Fe over the sum of the GEDR and GMDR models. This enhancement has 
been found in Oslo-type experiments and is confirmed independently by two-step 
cascade (TSC) measurements. To our knowledge, there exists at present no 
theoretical model which can explain an enhancement of this magnitude.

The first experiment, the $^{57}$Fe$(^3$He,$^3$He$^\prime\gamma)^{57}$Fe and 
$^{57}$Fe$(^3$He,$\alpha\gamma)^{56}$Fe reactions, was carried out with 45-MeV 
$^3$He ions at the OCL\@. Particle-$\gamma$ coincidences were measured by eight
Si particle telescopes at 45$^\circ$ and by an array of 28 NaI(Tl) 
$5^{\prime\prime}\times 5^{\prime\prime}$ $\gamma$ detectors with a solid-angle
coverage of $\sim$15\% of $4\pi$. The reaction spin window was 
$I\sim 2$--$6\ \hbar$. The 3.4-mg/cm$^2$-thick, self-supporting $^{57}$Fe 
target was enriched to $\sim 95$\%. The experiment ran for one week with a beam
current of $\sim 2$~nA\@. Total $\gamma$-cascade spectra were constructed in 
240-keV excitation-energy bins in the residual nuclei. These spectra were 
unfolded and a primary-$\gamma$ matrix $P$ was obtained by a subtraction method
\cite{GT96+GR87}. This matrix was factorized into a level density and total RSF
(summed over all multipolarities) according to the Brink-Axel hypothesis 
\cite{Br55+Ax62} by
\begin{equation}
P(E,E_\gamma)\propto\rho(E-E_\gamma)\,f_\Sigma(E_\gamma)\,E_\gamma^3.
\label{eq:ba}
\end{equation}
More details on the data analysis, including the normalized level densities of
$^{56,57}$Fe, are given in \cite{SA03} and references therein. 

\begin{figure}
\includegraphics[totalheight=4.3cm]{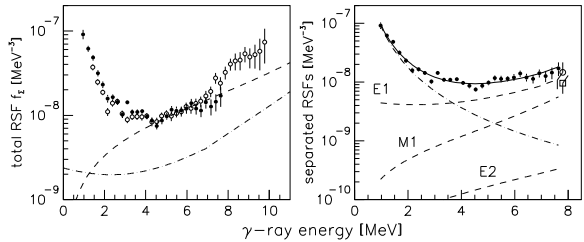}
\caption{Left panel: total RSF $f_\Sigma$ of $^{57,56}$Fe (filled and open 
circles, respectively), Lorentzian (dashed line) and KMF model (dash-dotted 
line) descriptions of the GEDR\@. Right panel: fit (solid line) to $^{57}$Fe 
data and decomposition into the renormalized $E1$ KMF model, Lorentzian $M1$ 
and $E2$ models (all dashed lines), and a power law to model the large 
enhancement for low energies (dash-dotted line). Open symbols are estimates of 
the $E1$ (circle) and $M1$ (square) RSF from hard primary-$\gamma$ rays 
\protect\cite{KU95}.}
\label{fig:fersf}
\end{figure}

RSFs are brought to an absolute scale by normalizing them to the average total 
radiative width $\langle\Gamma_\gamma\rangle$ of neutron resonances 
\cite{VG01}. There, the assumption of equal amounts of positive and negative 
parity states at any energy below $S_{n}$ is made. The violation of this 
assumption for low excitation energies introduces a systematic error to the 
absolute normalization in the order of $\sim 4\%$. In the case of $^{56}$Fe, 
also the value of $\langle\Gamma_\gamma\rangle$ has to be estimated from 
systematics. However, branching ratios needed for the subsequent analysis of 
TSC measurements are independent of the absolute normalization of the total RSF
and are consequently not affected by the above assumptions. The normalized RSFs
in $^{56,57}$Fe are displayed in Fig.\ \ref{fig:fersf}. The striking feature of
the RSFs is a large strength for soft transitions which has not been observed 
in the case of rare-earth nuclei, where we used the same analysis tools 
\cite{VG01}.

The soft transition strength constitutes a more than a factor of ten 
enhancement over common RSF models recommended in compilations \cite{Ob98}. To 
our knowledge, no other model can at present reproduce the shape of the total 
RSF either. A schematic temperature dependence of the RSF is taken into account
in the KMF model. It is, however, insufficient to describe the data. 
Phenomenologically, the data are well described as a sum of a renormalized KMF 
model, Lorentzian descriptions of the GMDR and the isoscalar $E2$ resonance, 
and a power law modeling the large enhancement at low energies
\begin{equation}
f_\Sigma=K\,(f_{E1}+f_{M1}+\frac{A}{3\,\pi^2\,c^2\,\hbar^2}\,E_\gamma^{-B})+
E_\gamma^2\,f_{E2}.
\label{eq:rsf}
\end{equation}
The parameters for the RSF models are taken from systematics \cite{Ob98}. The 
fit parameters for $^{57}$Fe are $K=2.1(2)$, $A=0.47(7)$~mb/MeV, and $B=2.3(2)$
($E_\gamma$ in MeV). However, the good description of the enhancement by a 
power law should not prevent possible interpretations as a low-lying resonance 
or a temperature-related effect.

To ensure that the observed enhancement is not connected to peculiarities of 
the nuclear reaction or analysis method, a TSC measurements based on thermal 
neutron capture has been performed to confirm the findings. It has been shown 
that TSC intensities from ordered spectra can be used to investigate the soft 
RSF \cite{VS03,BK95}. The TSC technique for thermal neutron capture has been 
described in \cite{BV91+BC95}. It is based on multiplicity-two events 
populating low-lying levels. Here, we will only give a brief description of 
some details.

The TSC experiment, i.e., the $^{56}$Fe$(n,2\gamma)^{57}$Fe reaction, was 
performed at the dual-use cold-neutron beam facility of the Budapest Research 
Reactor (see \cite{EB02,BR03} and references therein). About $2$~g of natural 
iron was irradiated with a thermal-equivalent flux of 
$3\times 10^7$~cm$^{-2}$s$^{-1}$ cold neutrons for $\sim 7$ days. Single and 
coincident $\gamma$ rays were registered by two Ge(HP) detectors of 60\% and 
13\% efficiency at a distance of 8~cm from the target. They were placed at 
$62.5^\circ$ with respect to the beam axis in order to minimize the effect of 
angular correlations. Chlorine and chromium targets as well as a certified 
$^{152}$Eu source have been measured to determine relative detector 
efficiencies up to 9~MeV $\gamma$ energy.

\begin{figure}
\includegraphics[totalheight=4.3cm]{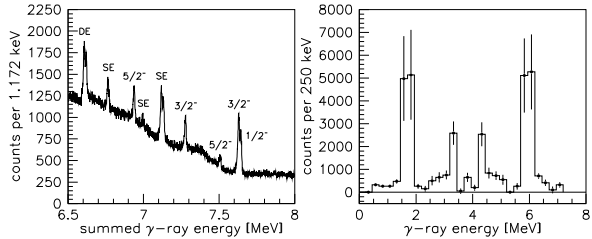}
\caption{Left panel: summed-energy spectrum. Peaks are labeled by the spin and 
parity of the final levels. SE and DE denote single- and double-escape peaks. 
Right: efficiency-corrected and background-subtracted TSC spectrum gated on the
unresolved doublet of the ground and first excited state. The spectrum is 
compressed into 250-keV-wide energy bins. Error bars include statistical errors
and intensity fluctuations within each bin.}
\label{fig:tsc}
\end{figure}

TSCs populating discrete low-lying levels in $^{57}$Fe produce peaks in the 
summed-energy spectrum shown on the left panel of Fig.\ \ref{fig:tsc}. Gating 
on the unresolved doublet of the $1/2^-$ ground state and the $3/2^-$ first 
excited state at 14~keV yields the TSC spectrum on the right panel of Fig.\ 
\ref{fig:tsc}. Spectra to other final levels were not investigated due to their
lower statistics and higher background. The TSC spectrum is compressed to 
250-keV-wide energy bins. It is symmetric around the mid point (half of the sum
energy) since always both $\gamma$ energies are recorded. When the sequence of 
the two $\gamma$ transitions is not determined experimentally, cascades with 
soft (discrete) secondary transitions are registered in the TSC spectrum as 
peaks on top of a continuum of cascades with soft primary transitions. Absolute
normalization of TSC spectra is achieved by normalizing to five strong, 
discrete TSCs for which absolute intensities of their hard primary transitions 
and branching ratios for their secondary transitions are known \cite{Bh98}. The
estimated error of the normalization is $\sim 20\%$; the method does not rely 
on the knowledge of absolute detector efficiencies. The statistical portion of 
the TSC spectrum, i.e., the remaining smooth part when peaks due to strong, 
discrete cascades are removed, usually exhibits a bell shape. However, in the 
present case, the smooth part has a rather flat shape, especially in the wings 
of the spectrum below $\sim 1.5$~MeV and above $\sim 6.2$~MeV\@. This is 
already an indication for an unusual enhancement of cascades with soft primary 
$\gamma$ rays. In the following, the smooth part of the TSC spectrum will be 
investigated in more detail.

\begin{figure}
\includegraphics[totalheight=4.3cm]{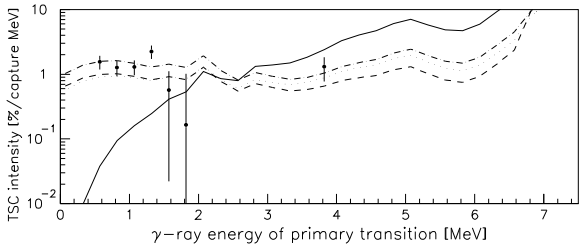}
\caption{Experimental TSC intensities (compressed to 250-keV-broad $\gamma$ 
energy bins) for cascades with soft primary $\gamma$ rays and at the mid point 
of the spectrum (data points with error bars). Lines are statistical-model 
calculations based on experimental data for the level density and $f_\Sigma$, 
neglecting (solid line) and assuming $E1$ (dashed line), $M1$ (dash-dotted 
line), and $E2$ (dotted line) multipolarity for the third term in Eq.\ 
(\protect\ref{eq:rsf}), i.e., the soft pole of the RSF\@.}
\label{fig:res}
\end{figure}

In order to separate cascades with soft primary and soft secondary transitions 
in the TSC spectra, we use the fact that the spacing of soft, discrete 
secondary transitions in regions of sufficiently low level density is 
considerably larger compared to the detector resolution. Thus, soft secondary 
transitions will reveal themselves as discrete peaks. On the other hand, soft 
primary transitions will populate levels which are spaced much closer than the 
detector resolution and will hence create a continuous contribution. Separation
of soft primary and secondary transitions is therefore reduced to a separation 
of individual peaks from a smooth continuum (by, e.g., a fitting procedure) in 
the appropriate energy interval \cite{BK95}.

The spin of the compound state in $^{57}$Fe populated by $s$-wave neutron 
capture is $1/2^+$. Thus, in the excitation-energy region 0.55--1.9~MeV, there 
are only three levels which can be populated by primary $E1$ transitions: the 
$1/2^-$ level at 1266~keV, the $3/2^-$ level at 1627~keV, and the $3/2^-$ level
at 1725~keV\@. All other levels have spins $5/2^-$ and higher and can only be 
populated by transitions with $M2/E3$ and higher multipolarity. Assuming that 
$\gamma$-transitions of such high multipolarities have a negligible 
contribution to the TSC spectrum, we do not take them into account in the 
further analysis. TSCs to the ground and first excited states involving the 
three above-mentioned levels as intermediate levels can easily be identified 
from their corresponding peaks in the TSC spectrum. Their contribution to the 
TSC spectra is subtracted. The remaining, continuous TSC spectrum in the 
specified energy range can be assigned to TSCs with soft primary 
$\gamma$-transitions. This smooth part of the TSC spectrum is used to test the 
soft RSF obtained from the Oslo-type experiment. Also, the mid point of the TSC
spectrum, where energies of primary and secondary transitions are equal (and 
hence, known) has been used in the subsequent analysis. For other energy 
intervals, the determination of the sequence of the two transitions in TSCs is 
subject to large uncertainties, thus, they are unsuitable for the present 
analysis.

In the present analysis, the intensity of ordered TSCs between an initial and 
final state is calculated on the basis of the statistical model of 
$\gamma$-decay from compound states
\begin{equation}
I_{if}(E_1,E_2)=\sum_{XL,XL^\prime,J_m^\pi}\frac{\Gamma_{im}^{XL}(E_1)}
{\Gamma_i}\rho(E_m,J_m^\pi)\frac{\Gamma_{mf}^{XL^\prime}(E_2)}{\Gamma_m},
\label{eq:tsc}
\end{equation}
where $E_1$ and $E_2$ are the energies of the first and second transition in 
the TSC which are connected by $E_i-E_f=E_1+E_2$. $\Gamma_{im}$ and 
$\Gamma_{mf}$ are partial and $\Gamma_{i}$ and $\Gamma_{m}$ are total decay 
widths of the initial and intermediate ($m$) levels, respectively. The average 
values of these widths can be calculated from the RSF by Eq.\ 
(\ref{eq:partial}). Summing in Eq.\ (\ref{eq:tsc}) is performed over all valid 
combinations of multipolarities $XL$ and $XL^\prime$ of transitions and of 
spins and parities of intermediate states. Thus, TSC spectra depend on the same
level density and RSFs which are extracted from the Oslo-type experiment, see, 
e.g., Eqs.\ (\ref{eq:ba},\ref{eq:rsf}).

Statistical-model calculations with experimental values for the level density 
and the total RSF have been performed assuming the decomposition of $f_\Sigma$ 
according to Eq.\ (\ref{eq:rsf}), and a standard spin-parity distribution for 
intermediate states \cite{GB03}. Four calculations were performed: one by 
neglecting the third term in Eq.\ (\ref{eq:rsf}), i.e., without the soft pole 
of the RSF, the other three under the assumption of $E1$, $M1$, and $E2$ 
multipolarity, respectively, for this term. In Fig.\ \ref{fig:res}, results are
compared to experimental data for energies where ordering of TSCs can be 
achieved. The calculation without the soft pole does not reproduce the data at 
all. The $\chi^2$ excluding the two data points at lowest $\gamma$ energies 
where we do not have experimental data on $f_\Sigma$ yields $\sim 25$, thus, 
ruling out this calculation on a statistical significance level higher than 
99.9\%. For calculations under the assumption of $E1$ and $E2$ multipolarities 
for the soft pole, the $\chi^2$ gives $\sim 9$. Thus, although these 
multipolarities cannot be ruled out on a significance level better than 
$\sim 85\%$, they are unlikely. The $\chi^2_{\text{red}}$ for the calculation 
with the $M1$ hypothesis for the soft pole equals $\sim 1.3$ and makes this the
preferred assignment. The selectivity between $M1$ and $E1/E2$ assignment 
becomes better when including the two data points at lowest $\gamma$ energy. 
However, there, the statistical-model calculation is based on an extrapolation 
of $f_\Sigma$ below experimental data, hence, possible systematic errors can 
become large. To check the sensitivity of our result, we have performed 
calculations with an exponential and resonance description of the enhanced soft
transition strength, both avoiding the pole for $E_\gamma\rightarrow 0$. The 
values of $\chi^2$ are rather insensitive to changes in the extrapolation of 
$f_\Sigma$. However, the experimental TSC intensities for the lowest two 
$\gamma$ energies are not so well reproduced as before. Finally, we have 
performed calculations where the ratio of the negative-parity levels to the 
total number of levels decreases linearly from $\sim 90$\% at 2.2~MeV to 
$\sim 50$\% at $\sim 7.6$~MeV excitation energy. As expected, TSC intensities 
with soft primary $\gamma$ rays are rather insensitive to this variation as 
well.

In conclusion, a more than a factor of ten enhancement of soft transition 
strengths (a soft pole) in the total RSF has been observed in Oslo-type 
experiments using the $^{57}$Fe$(^3$He,$\alpha\gamma)^{56}$Fe and 
$^{57}$Fe$(^3$He,$^3$He$^\prime\gamma)^{57}$Fe reactions. This enhancement 
cannot be explained by any present theoretical model. The total RSF has been 
decomposed into a KMF model for $E1$ radiation, Lorentzian models for $M1$ and 
$E2$ radiation, and a power law to model the soft pole. In a second experiment,
TSC intensities from the $^{56}$Fe$(n,2\gamma)^{57}$Fe reaction were measured. 
Statistical-model calculations based on separated RSFs from the decomposition 
of the experimental total RSF and on experimental level densities from the 
Oslo-type experiment were performed. These calculations can reproduce the 
experimental TSC intensities with soft primary $\gamma$ rays only in the 
presence of the soft pole in the total RSF\@. $M1$ assignment for the soft pole
is preferred, but $E1$ and $E2$ multipolarity cannot be ruled out on a 
significance level better than $\sim 85\%$. The satisfying reproduction of the 
experimental TSC data constitutes support for the physical reality of the soft
pole, independent from the Oslo-type experiment. It should be noted that this 
support was gained by using a different nuclear reaction, a different type of 
detector, and a different analysis method. Finally, as further supporting 
evidence, we would like to mention that preliminary results on a chain of 
stable Mo isotopes also indicate the presence of a soft pole in the total RSF
\cite{Ch04}, while in the case of $^{27,28}$Si, the Oslo method was able to 
reproduce the total RSF constructed from literature data on energies, 
lifetimes, and branching ratios available for the complete level schemes 
\cite{GM03}. 

Part of this work was performed under the auspices of the U.S. Department of 
Energy by the University of California, Lawrence Livermore National Laboratory 
under Contract W-7405-ENG-48. Financial support from the Norwegian Research 
Council (NFR) is gratefully acknowledged. Part of this work was supported by 
the EU5 Framework Programme under Contract HPRI-CT-1999-00099. G.M. and U.A. 
acknowledge support by U.S. Department of Energy Grant No.\ DE-FG02-97-ER41042.
Part of this research was sponsored by the National Nuclear Security 
Administration under the Stewardship Science Academic Alliances program through
DOE Research Grant No.\ DE-FG03-03-NA00076. We thank Gail F. Eaton and Timothy 
P. Rose for making the targets.

\end{document}